# Urban Scaling and Effects of Municipal Boundaries

Pieter P. Tordoir[1] and Anthony F.J. van Raan[2]
1. Department of Human Geography, Planning and International Development Studies, University of Amsterdam
   p.p.tordoir@uva.nl
2. Centre for Science and Technology Studies, Leiden University
   vanraan@cwts.leidenuniv.nl

*Abstract*

*Urban scaling, the superlinear increase of social and economic measures with increasing population, is an ubiquitous and well-researched phenomenon. This article is focused on socio-economic performance scaling, which could possibly be driven by increasing returns of the spatial size and density of interaction networks. If this is indeed the case, we should also find that spatial barriers to interaction affect scaling and cause local performance deviations. Possible barring effects of municipal boundaries are particularly interesting from the perspective of urban area governance policy and regional cooperation. To our best knowledge, this is the first study on this politically relevant and strongly disputed subject. We test the hypothesis of possible barring effects of municipal boundaries by correlating municipal boundaries with the structure of commuter networks within a large densely urbanized region, the Randstad in The Netherlands. The measured network impacts of these boundaries are subsequently correlated with local employment deviations. In order to spatially pinpoint correlations, we apply advanced spatially weighted modelling technique. We find that municipal borders have significant effects on inter-municipal commuting. Moreover we can indicate the specific effect of individual municipal borders precisely on the map. The results show particularly significant correlations along dividing lines between large urban agglomerations and rural communities. The southern part of the Randstad is more fragmented by such dividing lines than the northern part, which could partly explain the diverging economic development between the two parts.*

## 1.    Introduction

Urban scaling models are mostly applied to a set of cities defined as urban areas, whereby the scaling phenomenon is described on the basis of a superlinear statistical relation between socioeconomic performance -particularly in terms of the gross urban product (GUP), productivity, or employment (e.g., number of jobs)- as dependent variables, and population as the explanatory, independent variable [1,2]. This *urban scaling* relation is described by a power-law dependence of, for instance, the gross urban product on population size:

$$G(N) = aN^\beta \tag{1}$$

where $G$ is the gross urban product[1] and $N$ the population size of a city. The exponent $\beta$ follows from the measurement; in most cases, values of the exponent are between 1.10 and 1.20. Other units of analysis are cities defined as municipalities and larger urban regions [3-6]. For instance recently van Raan [4] discussed the superlinear increase of the gross urban product , productivity and employment with an increasing population of municipalities and regions in Germany, Denmark and the Netherlands. The GUP scaling exponent in the Netherlands is around 1.15 which means that a city twice as large (in population) as another city can be expected to have approximately a $2^{1.15}$=2.22 larger socio-economic performance (in terms of the gross urban product).

---
[1] Throughout the text of this paper we use the abbreviation GUP for the gross urban product. In the case of Eq.1 we use the shorter symbol *G*.

This exponent is in line with urban scaling findings across the globe. According to a widely held hypothesis in the model-oriented literature, the scaling of urban economic performance is driven by innovation-inducing social and economic interactions that increase disproportionally (superlinearly) with urban population size [7-10]. These growth-inducing interactions are known among economists and economic geographers as 'Marshallian externalities' and 'Jacobs externalities' [11]. Other, and additional explanations for urban scaling are provided by urban hierarchy and central places theory, and market gravitation [12,13]. In any of these explanations, face-to-face interactions are at center stage. We thus would expect that local physical, socio-cultural, and administrative interaction barriers diminish urban performance scaling or, in other words, that these barriers induce negative local performance residuals of the general scaling model. In this study, we investigate the role of municipal boundaries in this regard.

To find effects of boundaries on inter-local interaction and economic performance scaling, the spatial units for measurement should be more detailed than the areas enclosed by these boundaries. Therefore, in this study we go into further detail and investigate how scaling works at the micro-level of local neighborhoods, how this micro-scaling relates to macro-scaling, and how micro-scaling can reveal the effects of municipal boundaries. Do adjacent neighborhoods divided by municipal borders show less interaction and more negative scaling residuals than adjacent neighborhoods within municipalities, controlling for other explanatory variables, notably social and economic composition, and travel distance? The study focuses in this respect on inter-local interaction by way of spatial commuting patterns, whereby local and urban employment is taken as a dependent variable.

Investigating micro-scaling however requires another approach as compared to macro-level in which urban areas, municipalities or regions are the unit of measurement. The reason is that at the local level, clusters of economic activity such as ports, airports, manufactural sites, business parks, offices and shopping centers, are often spatially separated from population neighborhoods. This renders a direct scaling model senseless at the geographical micro-level. Only some fraction of the population living in a specific neighborhood will depend on the economy in that neighborhood; many people will commute to other areas in the wider environment. The essence of the urban scaling phenomenon is the actual spatial interaction field: the market area for contacts, exchanges of labor, exchange of ideas, etcetera. In theory and practice, it is therefore the spatially extended *daily urban system* that acts as the basis for the scaling mechanism. A model for urban scaling at the micro-level has to take this into account. In this study we therefore apply a spatial gravity function to micro-location employment data [14]. Micro-locations are the around 1,500 4-digit postal-code areas in the Randstad, the densely populated western part of the Netherlands with around seven million inhabitants, consisting of 92 municipalities among which the four largest cities in the country.

The structure of this paper is as follows. In Section 2 we discuss urban scaling at the macro level followed by our approach to analyze micro-level scaling. Next, we compare our empirical findings at the micro-level with those at the macro-level. Section 3 addresses a cornerstone of the study, the effect of municipal borders on daily commuting patterns. With this we answer the question where municipal borders hamper commuting and efficient functioning of the labor market. Subsequently, local border effects are correlated on their turn with local employment residuals of the micro-level scaling. We conclude our paper in Section 4 with a discussion of the policy-relevant consequences of our investigations.

## 2. Urban Scaling: Macro-Level Compared with Micro-Level

### 2.1 Macro-Level Data: Municipalities

A recently published study [4] discusses the scaling of municipalities in Germany, Denmark and the Netherlands, with gross urban (municipal) product (GUP) as the main variable[2]. Data on employment and productivity are also available in that macro-level study. Given our focus on the most densely populated part of the Netherlands, the Randstad, we use these data to analyze the urban scaling in this region. In our macro-level approach we define the Randstad as the province of South Holland (including the urban areas of Rotterdam, The Hague, Leiden and Dordrecht), plus the urban areas of Amsterdam, Utrecht and Haarlem, in total about 7 million inhabitants in 92 municipalities. In Fig. 1 we show the scaling of employment (number of jobs, Njobs) for all Randstad municipalities. The scaling exponent is 1.11 (95%CI [1.09-1.13], $R^2$=0.93). Next to the scaling exponent the deviations of the individual municipalities from the regression line are important. These scaling *residuals* are a measure of the difference between the actual employment of a municipality and the expected value based on the scaling exponent determined by the entire set. For the calculation of residuals we refer to [4]. We present the residuals of all Randstad municipalities in the Supplementary Information section, Table S1. If we look at the four largest cities (in order of population size: Amsterdam, Rotterdam, The Hague, Utrecht) we see the better position of Amsterdam and Utrecht as compared to Rotterdam and The Hague.

---

[2] The data are from LISA, the database of business locations in the Netherlands https://www.lisa.nl/home. The LISA data are based on location level which means one specific address of a company or institution. The number of employed persons is registered annually at that location, broken down by gender, full-time/part-time (hour limit 12 hours per week). Thus, all data is collected and registered per location, and not at a higher aggregation level, as is customary at the Central Bureau of Statistics (CBS). The LISA database consists of 1,630,070 locations (2018). All higher level data such as at the municipal level are aggregations of these location level data.

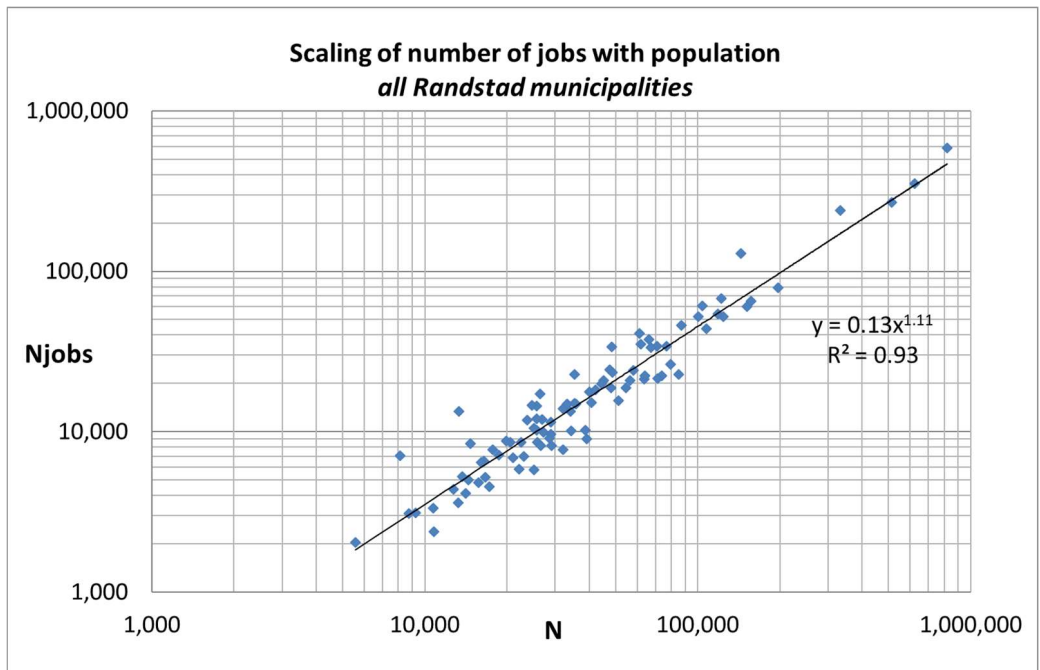

*Fig.1 Scaling of employment for all municipalities in the Randstad, 2014-2016. Njobs: number of jobs; N: number of inhabitants.*

In Table S1 (and also clearly visible in Fig.1) we see outliers, particularly municipalities with an exceptionally high residual. We take two examples at the low end (left side) of the population scale. Ouder-Amstel is a small Amsterdam suburb (population about 14,000), but it is the location of a large international bus station (Eurolines) combined with a major Amsterdam railway station with together about four million passengers per year. This clearly provides Ouder-Amstel many more jobs than a municipality of similar size. The second example is Zoeterwoude, a small Leiden suburb of around 9,000 inhabitants. This municipality is home to the large Heineken beer factory which gives this municipality thousands of jobs more than can be expected on the basis of its size. At the high end (right side) of the population scale we also see an outlier: Haarlemmermeer, a major Amsterdam suburb of about 150,000 inhabitants. The high positive residual of Haarlemmermeer can be explained very well: Amsterdam International Airport Schiphol, the fourth largest airport in Europe, is located in the municipality of Haarlemmermeer.

### 2.2 Micro-Level Data: Postal-Code Areas

There are no data on productivity and gross urban/municipal product (GUP) available at the micro-level. But we do have reliable data on employment and population at this level. This enables comparison with the findings at the macro-level. Thus, we take local employment as the dependent variable and population as the explanatory variable[3]. Units of measurement for the micro-scaling model are 1,500 postal-code areas in the Randstad region.

Scaling emerges from networked interactions between social and economic actors and population. This means that analysis at the micro level must be based on the spatial fields within which these interactions occur. Therefore, local employment and population are measured on the basis of their *gravity* values within a wider market system, indicating local market potential (for a discussion of urban gravity models see [13]). Employment and population of postal codes are increased by employment and population in adjacent postal codes, weighted by a linear and negative (car) travel time decay function, within a 45 minutes travel time circle. The real employment and population variables are thus transformed into gravity- or market potential variables, that account for both the centrality of locations within, and the size of the surrounding market field (the 'daily urban system' within 45 minutes travel distance, by car). The scaling model itself is a straightforward regression. The model fit shows that the scaling exponent is 1.12 (99%CI [1.11-1.12], $R^2$=0.98) and this value lies within the range that is often found in scaling studies. Because of the large number of units measured and the high correlation, the model is statistically very significant.

---

[3] Data source: Central Bureau of Statistics Netherlands (CBS): microdata SSB (Social Statistical Databases) and ABR (General Business Register) 2016. See https://www.cbs.nl/en-gb/our-services/customised-services-microdata/microdata-conducting-your-own-research/microdatabestanden/abr-general-business-register.

In the previous section we discussed the importance of residuals of individual units next to the scaling exponent. The scaling residuals in our micro-level approach are a measure of the difference between the actual (gravitated) employment of a postal-code area and the expected value based on the scaling exponent which is determined by the entire set. These residuals are calculated by the model algorithm, denominated as residual per (gravitated) local population unit, and mapped with GIS (Choropleth mapping). The results of the residuals analysis are presented in Fig. 2, left panel. Negative employment residuals are indicated in red and positive residuals in blue, so that spatial patterns are easily discernable.

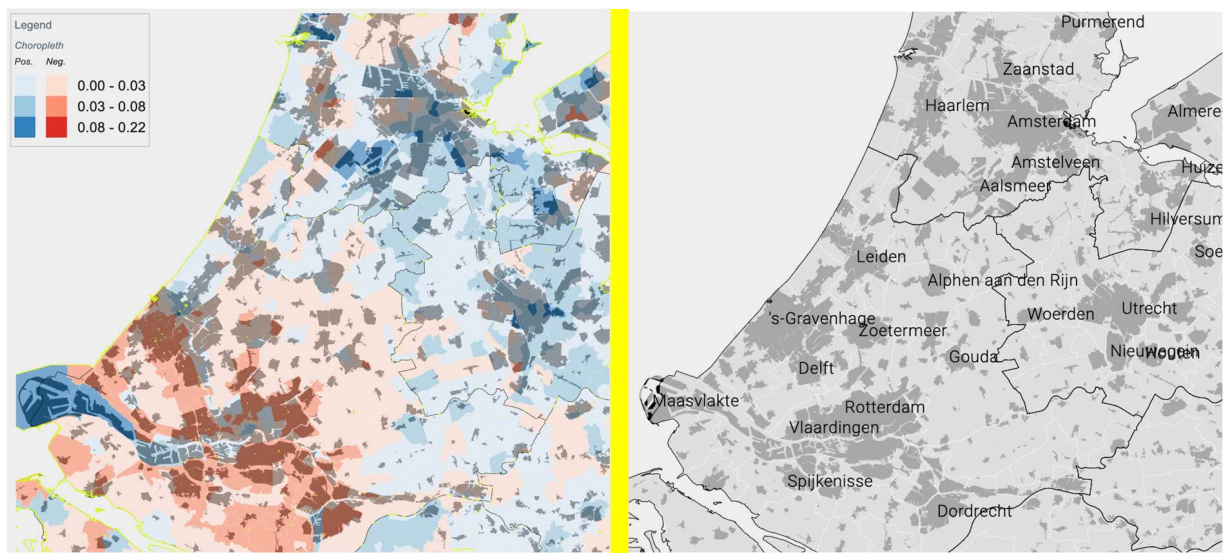

*Fig.2 Left panel: residual values of micro-scaling of employment in the Randstad region (residual employment per inhabitant, 2017). Blue colors indicate positive residuals, red colors negative residuals. Right panel: names of the major cities in the region. ('s Gravenhage = The Hague).*

The map of residuals shows a clear pattern. The northeastern part of the Randstad region, known as the *North Wing* with Amsterdam and Utrecht as core cities, is dominated by blue, which means positive employment residuals. The *South Wing* with the core cities of Rotterdam and The Hague is dominated by red, negative residuals. The Rotterdam Port Area is one of the few exceptions in this general pattern. This economic lagging of the South Wing vis-à-vis the North Wing is a well-known fact in The Netherlands. Employment growth in the predominantly blue collar sectors in the Rotterdam region and government services in The Hague region has been much slower compared to the advanced commercial services and high-tech activities that characterize the North Wing economy.

The scaling exponent in this micro-analysis of 1.12 for the Randstad lies well within the error margin of the scaling exponent 1.11 measured with the macro-level data for the Randstad discussed in the previous section. We see that the difference between macro-level scaling exponent based on entire municipalities and the micro-scaling exponent of postal-code areas within the same region is negligible. It is clear that micro-scaling based on 'gravitated' postal-code areas and macro-scaling based on municipalities give comparable results. Therefore we feel confident to use the micro-level data for an investigation of an economically, socially and politically important issue: do municipal boundaries have an effect on the socioeconomic development and, with that, on the welfare of the municipalities involved and, in the end, of the whole country?

## 3. Effects of Municipal Boundaries on Micro-Scaling Residuals

### 3.1 First analytical step: effect of boundaries on commuting interaction

In order to investigate and explain the effects of municipal boundaries on scaling residuals, we take two consecutive steps. First, we investigate with a spatial interaction model the actual interactions in terms of the daily flows of workers from residential areas of origin to work destination areas, again at the micro level of postal-code areas[4]. This enables us to detect barring effects of municipal boundaries. In the next step we measure with a spatial regression model where and to which degree the boundary effects found in the first step explain the micro-level scaling residuals discussed in the previous section.

In order to measure where municipal boundaries influence commuter flows, we apply a *spatially weighted interaction model* (SWIM) on an origin-destination matrix of 1,500 postal-code areas in the Randstad region. The model is filled with microdata of the residential address (origin) and workplace address (destination) of all employed and tax-paying citizens in the Randstad region. The SWIM model is an elaboration of the gravitational spatial interaction model [14, 15], whereby parameters of individual locations such as postal codes acquire fitted coefficients on the basis of their weighted position in surrounding subfields (or *kernels*) of postal codes, for each of which the interaction model is fitted in a series of runs. Mathematically, the basis is a spatial (gravity) interaction model:

$$T_{ij} = \kappa \, R_i^\alpha \, W_j^\gamma \, d_{ij}^\beta \qquad (2)$$

whereby $T_{ij}$ is the estimated commuting between areas $i$ and $j$; $R_i$ represents the residential employed population in area $i$; $W_j$ is the number of individual working places in area $j$; $d_{ij}$ is the travel time (by car) between $i$ and $j$; and $\kappa$, $\alpha$, $\gamma$, and $\beta$ are the parameters to be estimated on the basis of the empirical origin-destination matrix, they describe the relationship between the spatial commuting flows and each of the explanatory

---

[4] Microdata about these commuter interactions are from the CBS in particular data concerning living and working addresses of employed persons (CBS-SBB 2017).

variables. In this first step (Eq.2) we have not yet included a specific municipal boundary value.

A first run of a commuter flows SWIM without a municipal boundary variable shows a total explained variance $R^2$=0.68 (99%CI: p=0.0001). We present in Fig. 3 the map of the residuals, where less-than-expected actual commuter flows, i.e., negative residuals are given as red lines; positive residuals are given as blue lines.

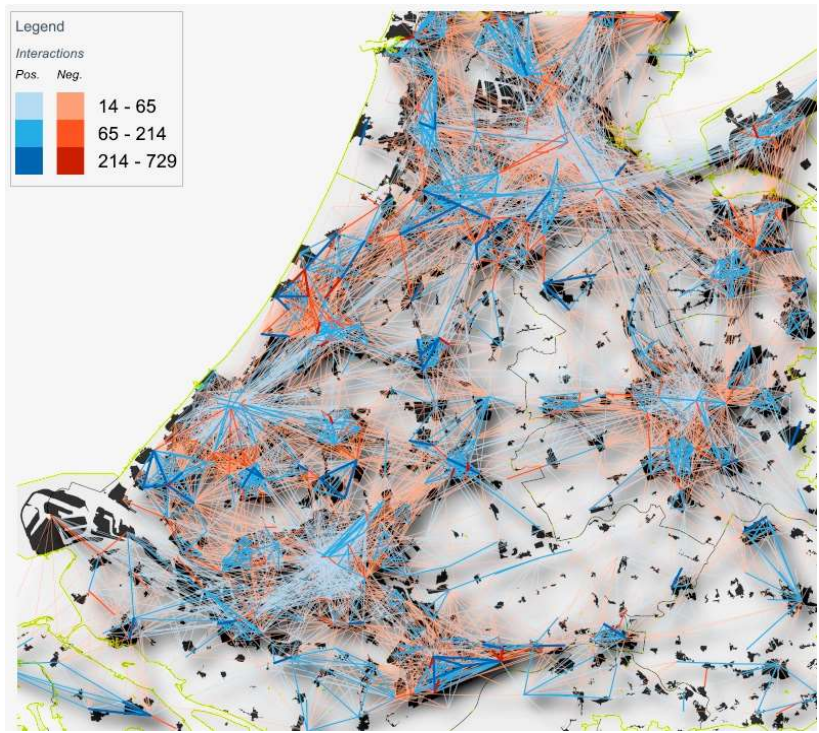

*Fig.3 Residual values of commuter flows in The Randstad region calculated with the spatially weighted model. The values are absolute numbers: more (blue lines) or less (red lines) observed commutes between postal-code areas than expected by the model (note that residual values in Fig. 2 are relative values).*

The above map contains dense information, from which four general patterns emerge. A first pattern consists of bundles of blue lines emanating from the center of Amsterdam to the centers of Utrecht, Leiden and The Hague. These actual commuter links, not expected by the model, comprise predominantly highly educated and young workers in advanced services industries, the 'urban gentry' living in one metropolitan center (center of a major urban area) and working in another metropolitan center. For this group the whole Randstad is one daily urban system. Rotterdam, the large port city, is however less connected in this respect. A second pattern shows strong commuter links between new towns and cities and specific work locations at relatively long distances: the larger-than-expected flows between the new city of Almere (in the upper right corner of the map, within the Amsterdam urban area) and Schiphol International Airport south of Amsterdam (city of Haarlemmermeer), and the flows between the new city Zoetermeer (within the The Hague urban area), and The Hague, the seat of the Dutch government. Thus in fact this second pattern concerns commuter links within major urban areas (Amsterdam and The Hague). A third pattern is composed of blue and therefore again more-than-expected commuter flows in and around medium-sized cities and rural areas. This denotes local and sub-regional clustering of specific industries and qualities in the local labor force. And finally, the fourth pattern is a strong overrepresentation of red lines, or less-then-expected commuting, between major cities and surrounding medium-sized cities and rural areas. Red lines are significant in the South Wing of the Randstad region, particularly between The Hague and directly surrounding suburbs (except for Zoetermeer), and between Leiden and its adjacent coastal suburbs.

In order to find municipal boundary effects on the empirical origin-destination matrix, we add in Eq.2 a variable $M$ for observed municipal boundary crossings of commutes:

$$T_{ij} = \kappa R_i^\alpha W_j^\gamma d_{ij}^\beta M_{ij}^\chi \quad (3)$$

The parameter $\chi$ describes the observed relationship between municipal boundary crossing $M$ and the size of commuter flows, analogous to the determination of the parameters for variables $R$, $W$ and $d$. As a final step, we apply spatially weighted regression to this basic model in order to find specific parameter values for each individual commuter origin area $i$. The resulting origin-focused Spatially Weighted Interaction Model (SWIM) takes a multitude of local calibration points $u$ (extensively discussed in [15]), each with surrounding kernels of 100 postal-code areas, for which individual regressions are fitted:

$$T_{ij} = \kappa_{\{ui\}} R_i^{\alpha_{\{u,i\}}} W_j^{\gamma_{\{u,i\}}} d_{ij}^{\beta_{\{u,i\}}} M_{ij}^{\chi_{\{u,i\}}} \quad (4)$$

This spatial weighting procedure renders the model much more precise, and it enables us to pinpoint specific local municipal boundary effects ($M$) on the map. The SWIM regression thus takes varying sub-regional circumstances into account, i.e., travel propensities and distance decay exponents that might vary with varying local urban density. This sub-regional specificity of SWIM regression boosts explanatory power.

The municipal boundary variable ($M$) is simply a source-directed dummy with value 0 (an individual commuter stays within the boundary of the municipality of residence) or 1 (the boundary is crossed). The addition of this variable renders clarity to what is actually measured: does the mere fact that a job is located *outside* the municipality of residence affect the propensity of an inhabitant of that municipality to take that job, controlling for the spatial distribution of jobs ($W_j^\gamma$) and the propensity to travel ($d_{ij}^\beta$)? Theoretically, these boundary effects, if they occur, can be expected to have a disruptive effect on a 'rational' interaction between the population and the economy. For instance: a citizen living near the municipal boundary chooses a specific location for working in his or her own municipality while there is across the municipal border a better job opportunity that would take less travel time and costs. The result is more travel time and less efficient labor markets, which will ultimately come at the expense of over-all employment and prosperity.

The fitted $M$-included model adds 13% to the variance explained by the $M$-excluded model; total variance explained rises to 82%, with significance levels for all sub-regional kernels well below p=0.01. Municipal borders therefore do have significant effects on inter-municipal commuting. Moreover, using the SWIM-technique, we can pinpoint the specific effect of individual municipal borders precisely on the map, see Fig. 4.

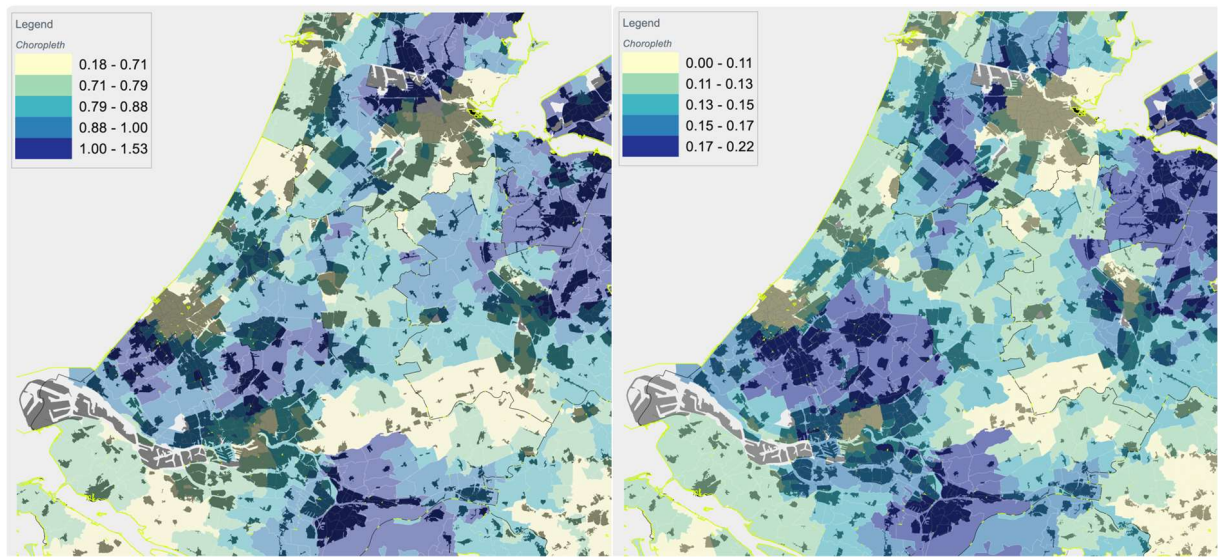

*Fig. 4. Municipal border effects on commuter flows in the Randstad, measured for postal-code areas. Left panel: observed local model parameter values of the effect. The higher the local parameter value, the stronger the dampening of inter-municipal commuting by local municipal borders. Right panel: total variance of commuter flows explained by local municipal borders ($R^2$). See the legend color codes for the strength of the municipal boundary parameter (left map) and the fraction of variance explained (right map). Source: CBS microdata SSB2017.*

The above maps give both the local effects of municipal borders on commuting, in terms of parameter values for the variable $M$ (left map) and the total variance of commuting that is locally explained by municipal borders (right map). In other words, the left map indicates the strength of the effect 'per se', whereas the right map indicates the weight of this effect relative to other explaining variables $R$, $W$, and $d$. Strength and variance are naturally interrelated, but not by 100%. In some locations we find relatively mild local effects that nonetheless explain commuting patterns better than distance decay $d$. Particularly in the Randstad South Wing, municipal boundaries explain much commuter flow variance, even in locations where the boundary effect is by itself less pronounced.

The maps show that municipal boundary effects on commuter flows are far from evenly distributed over the Randstad region. The centers of the largest cities are relatively free from these effects, as are rural areas at some distance from large cities. Strong effects however are encountered in relatively densely populated areas close to the largest cities, many of these being large suburban developments located within or adjacent to 'native' rural communities. The boundary effects relate in particular to commutes over relatively short distances, mostly between urban locations on one side of a municipal border and rural communities on the other side. Urban-rural commutes are in many places much less common than expected by the fitted model (see the red lines in Fig. 3). This lifts the veil somewhat regarding possible causes of boundary effects.

This study measures whether there are effects, but it does not immediately explain what causes them. Nonetheless, the measured geographical distribution of the effects feeds hypotheses about possible causes. Theoretically, we could expect strong boundary effects in case of wide social, economic, cultural and political differences between areas on different sides of a municipal border. These differences could very well hamper social and economic interaction across the border. A divide between on the one hand typically 'urban' populations and culture, and on the other hand typically 'rural' populations and culture, is well documented in the sociological and geographical research. These kinds of divisions could very well be prominent in the deep blue areas in the above maps. Diverging policies and a lack of governance coordination and integration across institutional boundaries can of course play a major role. Because the analysis indicates where effects play geographically, and with the knowledge of local issues and characteristics in mind, hypotheses about actual causes can be drawn up for testing in further research.

## 3.2 Second analytical step: relating boundary effects to employment scaling residuals

Now that municipal boundary effects on spatial interaction (commuter flows) are localized, we can measure to which degree these local effects are related to local employment. On the theoretical assumption that spatial interactions—among which commutes—drive urban scaling, we would expect local municipal boundary effects on interaction to be negatively related to local employment. Controlling for local population size and thus for urban scaling, boundary effects could in this theory explain negative employment scaling residuals that we showed in Fig. 2. However, there is an important caveat to this theoretical assumption. The assumption is basically about the efficient functioning of markets; labor markets in our case. Free movement of a large number of participants is prerequisite for efficiency in market theory. In reality, however, a closed 'club' or 'clan' of actors can also generate positive economic outcomes, under specific circumstances even better than markets can do [16]. For example, small and relatively isolated towns or regions can specialize in trades by making use of strong social and family ties. Such economically vibrant local communities abound within the Randstad region. It could be that urban jobseekers, who do not have access to these closed local communities, bear the brunt, and not the local community members. Thus, the relation between local border effects and employment, or any other economic indicator, could in fact be rather complex, with either negative or positive directions depending on the social embeddedness of the local economy [17].

This expected complexity can definitely be found in the Randstad region. A simple and spatially *un*weighted regression of municipal boundary effects and employment scaling residuals for the Randstad region as a whole does for that reason not render significant results. However, things change when we apply a spatially weighted regression and take thus local and sub-regional particularities into account. In many areas there is no significant relation between boundary effects and employment residuals, but in specific areas this relation is certainly present, see Fig. 5.

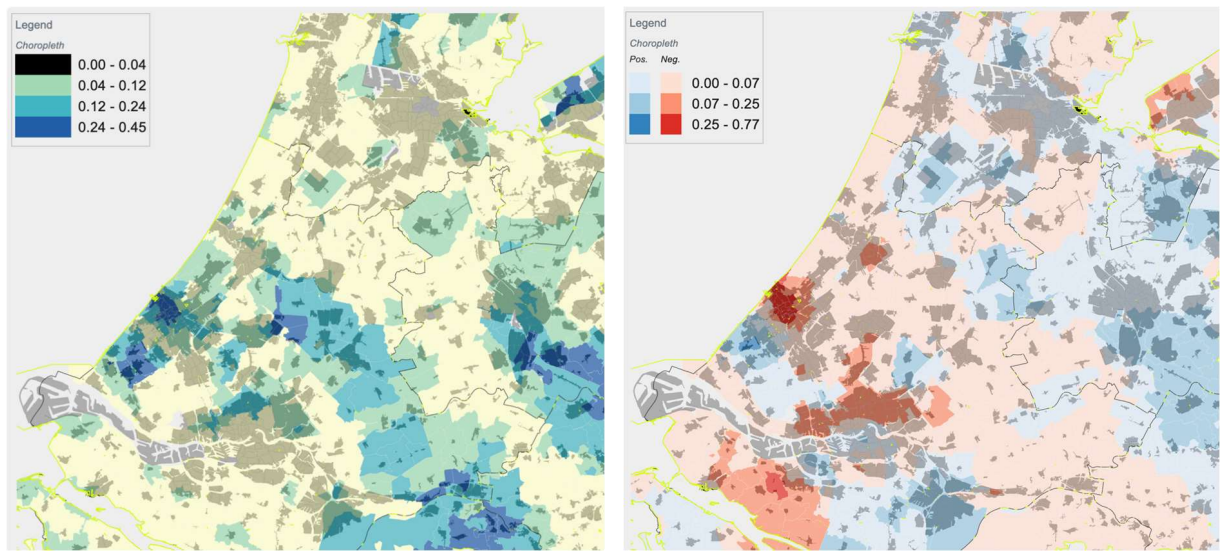

*Fig.5 Local employment scaling residuals explained by local municipal boundary effects on commuter flows (spatially weighted regression of boundary parameter values). Left panel: locally explained variance of the scaling residuals ($R^2$). Values 0.00-0.04 are not significant, hence the black color code. Right panel: specific coefficients of the local relation. Red areas: border effects act as barriers and push local employment downwards. Blue areas: border effects act as shields and push local employment upwards.*

We show the effects of municipal boundaries on local scaling residuals in two maps, whereby the left map of Fig. 5 indicates which specific boundaries have an effect on local employment, and the right map of Fig. 5 indicates whether this effect is negative (red areas) or positive (blue areas). The distinction between the two maps is relevant as we explained earlier: a municipal boundary effect could work beneficial for a specific area, but detrimental for surrounding areas. We see this, for example, strongly in the case of the municipal border between the city of The Hague and the Westland municipality that lies southwest of The Hague.

The above maps indicate significant negative municipal border effects on local employment (95%CI) in specific areas, that are more common in the South Wing than in the North Wing. The right map shows that the cities of The Hague, Rotterdam and Leiden suffer from negative employment effects; the Westland community adjacent to The Hague (southwest of that city) enjoys however a positive employment effect of its municipal borders. Westland is known for its world-class greenhouse horticulture but also for its religiosity, and tight community bonding. The border with The Hague, much less crossed by commuters than is predicted by our model (see Fig. 2), works in fact as a barrier for that city. Seen from the other side, the border functions as shield for the Westlanders. In general, the South Wing is socially, economically and culturally more fragmented than the North Wing and the Amsterdam urban area in particular. This fragmentation is partly reflected in municipal boundaries; boundaries might on their turn also reinforce social and economic fragmentation. The fragmentation is by and large detrimental for the large cities, but not necessarily so for rural communities. But given the importance of the cities this fragmentation in the South Wing could explain at least partly the lagging employment of the South Wing as a whole.

The North Wing certainly contains areas where municipal boundaries correlate with negative commuting residuals, but only in the city of Almere, a new large suburb of Amsterdam with about 215,000 inhabitants, located in the upper east corner of the Randstad map, do boundary effects correlate with negative local employment residuals. Other cities and towns in the east side of the Amsterdam urban area and north of Utrecht show boundary effects on inter-local interaction, but since these suburban cities disproportionally accommodate prosperous and well-educated residents who mainly commute to Amsterdam and Utrecht, this inter-local fragmentation will have less impact on the local economy. More in general, commuter trips are on average longer in the North Wing than is the case in the South Wing, which indicates wider interaction fields and benefits of agglomeration. Local boundary fragmentation might have less economic impact in those circumstances. The South Wing has a more complex and polycentric urban structure. Inter-local fragmentation might in these conditions have a relatively strong economic impact.

## 4. Conclusions and discussion

Urban population scaling of economic performance, including employment, is caused by face-to-face interaction. In advanced and urbanized services economies, activities spring from and settle in nodes of interaction networks. The wider the 'feeding' area for these nodes, the more they can thrive in a market economy. Barriers for interaction are detrimental to this mechanism, but more isolated areas could draw on tight local bonds to bypass a malfunctioning market. In this study we investigated in this respect the impact of municipal boundaries on interaction networks (commuting) and, via this, local economic performance (employment), controlled for general population scaling. To our best knowledge, this is the first study on this politically relevant and strongly disputed subject. To find boundary effects, we descended to the local micro-level and applied an adapted measurement technique, gravitated recalculation, that proved to render reliable results.

In the case of the densely populated Randstad region in The Netherlands, we found that municipal boundaries in specific areas are indeed related with local deviations from the general line of urban employment scaling. Large cities in particular can suffer a negative deviation; surrounding rural communities sometimes win. Particularly at intersections of urban and more rural areas, municipal boundaries act as a brake on labor mobility and the matching of supply and demand of talent and skills. Potential economies of scale and synergy are thus not converted into prosperity. As a result, the population in the central cities does not benefit enough from activity growth in nearby smaller centers, and smaller centers do not benefit enough from central city functions.

These institutional boundary effects far from explain all employment deviations; other factors play as well. We did not investigate reasons behind border effects but we can surmise that social, economic and cultural divides between municipal communities play a role. Such divides could of course also cut *through* municipalities, causing local employment deviations that relate less with municipal boundaries. With the technique of spatially weighted regression fitting microdata, it is possible to test this hypothesis in further research.

On a more general level we see a striking coincidence between employment underperformance and pronounced negative effects of municipal boundaries in the Randstad South Wing, a coincidence that is lacking in the North Wing. The South Wing, encompassing the Province of South-Holland, with Rotterdam (urban area population 1,3 million) and The Hague (urban area population 1,1 million) as largest cities, followed by Leiden (urban area 0.4 million) and Dordrecht (urban area 0.3 million), contributes more than 20% to the national total in terms of export, innovation and employment. The region suffers from structural problems; competing urban areas in Europe such as Antwerp and Munich have grown more than three times since 2009 [18]. This study, part of a research program on the relation between governance structure and regional economic development in the province of South Holland [19, 20], indicates that potential economic agglomeration benefits are not being sufficiently exploited, while they could be an important source of further prosperity. Our observation that administrative boundaries influence employment negatively in South Holland is in this respect a relevant research outcome for the practical organization of local and regional governance, with the aim of economic prosperity and competitiveness.

The global transition of the economy -characterized by, particularly, knowledge intensification, ICT and big data, digitization and robotization -does not automatically lead to an economy in which accessibility and proximity are irrelevant. On the contrary. Agglomeration, regional and spatial cohesion and synergy in the urban context are important for economic competitiveness and prosperity. Understanding the foundations for this spatial cohesion, and the stimulating or hindering role of administrative boundaries, is crucial for policymaking. Are spatial economic facilities and conditions such as business parks, science parks, accessibility networks, efficient and accessible labor market, and knowledge center development, picked up at an optimal spatial scale? Is optimal use made of potential agglomeration benefits and do administrative structures still match with the contemporary dynamics of the spatial economy? Our conclusion from the in-depth analysis of the province of South-Holland is: no. It is logical and inevitable that economic development, interactions, and synergy extend beyond municipal boundaries. As the Netherlands Environmental Assessment Agency points out in a recent report: "the economy functions on a larger scale than the administrative boundaries of the city" [21]. Consistency and synergy of governance in an expanding urban area context is becoming crucially important.

# Supplementary information

*Table S1. Residuals of the scaling of the number of jobs for the municipalities in the Randstad (outliers, as discussed in the main text, are in italics).*

| Major urban areas | Municipalities in the urban area | residual |
|---|---|---|
| **Amsterdam** | Amsterdam | 0.23 |
| | Zaanstad | -0.17 |
| | Amstelveen | 0.17 |
| | Diemen | 0.51 |
| | Wormerland | -0.19 |
| | Landsmeer | -0.14 |
| | Oostzaan | -0.03 |
| | Edam-Volendam | 0.05 |
| | Almere | -0.19 |
| | *Haarlemmermeer* | *0.64* |
| | Purmerend | -0.29 |
| | Waterland | -0.34 |
| | *Ouder-Amstel* | *1.02* |
| | Gooise Meren | -0.14 |
| | | |
| **Rotterdam** | Rotterdam | 0.02 |
| | Capelle a.d. IJssel | 0.27 |
| | Schiedam | 0.02 |
| | Barendrecht | 0.21 |
| | Lansingerland | -0.03 |
| | Nissewaard | -0.50 |
| | Vlaardingen | -0.37 |
| | Ridderkerk | 0.11 |
| | Albrandswaard | 0.08 |
| | Hellevoetsluis | -0.43 |
| | Krimpen a.d. IJssel | -0.17 |
| | Maassluis | -0.51 |
| | Brielle | 0.07 |
| | Westvoorne | -0.22 |
| | | |
| **The Hague** | The Hague | -0.03 |
| | Westland | 0.26 |
| | Delft | 0.14 |
| | Zoetermeer | -0.09 |
| | Rijswijk | 0.52 |
| | Leidschendam-Voorburg | -0.37 |
| | Pijnacker-Nootdorp | -0.33 |
| | Wassenaar | 0.02 |
| | Midden-Delfland | 0.02 |
| | | |
| **Utrecht** | Utrecht | 0.34 |
| | Stichtse Vecht | -0.21 |
| | Nieuwegein | 0.45 |
| | IJsselstein | -0.03 |
| | Zeist | 0.28 |
| | Houten | 0.14 |
| | De Bilt | 0.05 |
| | Bunnik | 0.45 |
| | | |
| **Leiden** | Leiden | 0.19 |
| | Leiderdorp | 0.12 |
| | *Zoeterwoude* | *0.93* |
| | Oegstgeest | -0.24 |
| | Voorschoten | -0.53 |
| | Katwijk | -0.25 |
| | Teylingen | 0.03 |
| | Noordwijk | 0.37 |
| | Lisse | -0.01 |
| | Hillegom | -0.15 |
| | Noordwijkerhout | 0.08 |
| | | |
| **Haarlem** | Haarlem | -0.13 |
| | Heemstede | -0.24 |
| | Bloemendaal | -0.37 |
| | Velsen | 0.14 |
| | Beverwijk | 0.07 |
| | Heemskerk | -0.57 |
| | Castricum | -0.31 |
| | Zandvoort | -0.17 |
| | Uitgeest | -0.29 |
| | Haarlemmerliede en Spaarnw. | 0.10 |
| | | |
| **Dordrecht** | Dordrecht | 0.00 |
| | Zwijndrecht | 0.08 |
| | Sliedrecht | 0.42 |
| | Papendrecht | 0.08 |
| | Alblasserdam | 0.16 |
| | Hendrik-Ido-Ambacht | -0.34 |
| | Hardinxveld-Giessendam | 0.15 |
| | | |
| **other cities/towns** | | |
| | Alphen aan den Rijn | -0.11 |
| | Nieuwkoop | -0.07 |
| | Kaag en Braassem | -0.16 |
| | Gouda | 0.10 |
| | Krimpenerwaard | -0.21 |
| | Zuidplas | -0.09 |
| | Bodegraven-Reeuwijk | 0.11 |
| | Waddinxveen | 0.19 |
| | Gorinchem | 0.47 |
| | Molenwaard | 0.00 |
| | Leerdam | 0.09 |
| | Zederik | 0.05 |
| | Giessenlanden | -0.06 |
| | Goeree-Overflakkee | -0.07 |
| | Oud-Beijerland | 0.25 |
| | Binnenmaas | -0.22 |
| | Cromstrijen | -0.05 |
| | Strijen | 0.03 |
| | Korendijk | -0.47 |